\documentclass[twocolumn,floatfix,pra,superscriptaddress,showpacs,eqsecnum]{revtex4}
\usepackage{amssymb}
\usepackage{amsmath}
\usepackage{graphicx}
\usepackage{bm}

\begin{document}

\title{Paired phases and Bose-Einstein condensation of spin-one bosons with attractive interaction}
\author{G. Pe\l ka}
\affiliation{Institute of Theoretical Physics, University of Warsaw, ul. Ho\.za 69, PL-00--681 Warszawa, Poland}
\author{K. Byczuk}
\affiliation{Institute of Theoretical Physics, University of Warsaw, ul. Ho\.za 69, PL-00--681 Warszawa, Poland}
\affiliation{Theoretical Physics III, Center for Electronic Correlations and Magnetism, Institute of Physics, University of Augsburg, D-86135 Augsburg, Germany}
\author{J. Tworzyd\l o}
\affiliation{Institute of Theoretical Physics, University of Warsaw, ul. Ho\.za 69, PL-00--681 Warszawa, Poland}

\date{\today}

\begin{abstract}

We analyze paired phases of cold bosonic atoms with the hyper spin $S=1$ and with an attractive interaction. We derive mean-field self-consistent equations for the matrix order parameter describing such paired bosons on an optical lattice. The possible solutions are classified according to their symmetries. In particular, we find that the self-consistent equations for the $SO(3)$ symmetric phase are of the same form as those for the scalar bosons with the attractive interaction. This singlet phase may exhibit either the BCS type pairing instability (BCS phase) or the BEC quasiparticle condensation together with the BCS type pairing (BEC phase) for an arbitrary attraction $U_0$ in the singlet channel of the two body interaction.  We show that both condensate phases become stable if a repulsion $U_2$ in the quintet channel is above a critical value, which depends on $U_0$ and other thermodynamic parameters.

\end{abstract}


\maketitle

\section{Introduction}
\label{intro}

Superconductivity in metals is a consequence of pairing between electrons \cite{Cooper56} and formation of a new macroscopic coherent state made of electron pairs \cite{Bardeen57}. The microscopic Bardeen-Cooper-Schrieffer (BCS) theory of superconductivity explains all key experimental features of the superconducting state. Recent developments in trapping, cooling, controlling, and detecting of atoms allowed to investigate superfluidity in neutral fermionic systems, cf. reviews:~\cite{Jak05,Blo08}. In particular, a crossover between Bose-Einstein condensation (BCS) limit, where fermionic pairs overlap significantly, and the BEC limit, where tightly bound  pairs form a coherent state of Bose-Einstein condensed bosons \cite{Leggett80,Noz85}, was demonstrated experimentally \cite{Reg03}. Since both fermionic as well as bosonic atoms are available in experiments it is natural now to investigate pairing between bosons and a coherent superfluid state of paired bosons.

Pairing and phase transitions of bosons with zero spin and with an attractive interaction was  discussed in Ref.~[\onlinecite{Eva65}] in the context of superfluid helium four. It was found that collective excitations of a coherent condensed state of paired bosons can undergo another BEC type condensation into a one-particle condensed state, now known as the Evans-Rashid transition \cite{Eva65,Eva73,Sto94}. However, already in Ref.~[\onlinecite{Sto94}] it was found that both the homogeneous coherent paired phase and the homogeneous phase due to the Evans-Rashid transition are unstable against a mechanical collapse. I.e., the bosons with an attractive interaction tend to form clusters of particles. Moreover, extending the mean-field result of [\onlinecite{Eva65,Eva73,Sto94}] by the leading-order fluctuation contributions \cite{Mue99,Jeo02} or higher order corrections \cite{Man10} to the thermodynamic potential due to the interaction between particles does not stabilize homogeneous phases of bosons with pairing potential. It is observed in \cite{Zin08} that by confining bosons with paring interaction in a trap, which produces a gap between the ground and the excited states, one can protect the system from the mechanical instability.  On the other hand, in Ref.~\cite{Sto08} a narrow region at finite temperatures on a phase diagram is found, where many-body effects can stabilize the homogeneous normal phase of  bosons with pairing interaction.

In the present paper we employ a mean-field theory to solve a problem of pairing between bosons with spin $S=1$ moving on an optical lattice. We show that the coherent BCS type phase of paired bosons induced by attractive interaction in the singlet channel is stable povided that the interaction in a quintet channel is repulsive and sufficienlty strong. Since bosons with  the nonzero hyperspin are available in experiments with cold atoms \cite{Jak05,Blo08} and  both sign and strength of the interaction between them can be tuned by the optical Feshbach resonance \cite{Tha05} it is interesting to look for such a paired bosonic state in a laboratory. 

The paper is organized as follows: In Section II we introduce the model. In Section III  spinor pair condensed phases are classified  by the matrix BCS type order parameter and the mean-field theory solution of the Hamiltonian is presented. Section IV is devoted to  the symmetry classification for the phases characterized by the complex matrix order parameter following  the standard approaches \cite{Vol86,Vol00,Yip07}. It should be noted that the finite spin of bosons with repulsive interaction in all channels \cite{Ohm98,Ho98} leads to many nontrivial ground and excited states of the spinor condensates \cite{Blo08,Ued10}. They include topologically nontrivial phases \cite{Koa00,Ued02,Yip07}, skyrmion excitations \cite{Kha01},  and even nonabelian vortices \cite{Sem07}. The symmetry classification presented in this Section may help in the future for detailed analysis of non-trivial excitations in paired phases of bosons. In Section V we present numerical solutions to the mean-field equations and in Section VI stability of particular phases is discussed. Conclusions are in Section VII and details on our derivations are presented in Appendices. 

\section{The model}
\label{model}

We consider the Hubbard model for spin-one bosons,  
which are trapped on an optical lattice. The grand canonical 
Hamiltonian \cite{Fis89,Oos01}
\begin{equation}
H=H^0+H^{\rm int}-\mu N\label{hubbard}
\end{equation}
contains the kinetic part $H^0$ and the interaction part $H^{\rm int}$.
We introduce the chemical potential $\mu$ to fix the average total number
of bosons in the lattice.
The kinetic part is 
\begin{equation}
H^0=-t\sum_{\langle i,j\rangle \sigma} b_{i\sigma}^{\dagger} b_{j\sigma}, \label{kinetic}
\end{equation}
where $b_{i\sigma}$ ($b_{i\sigma}^{\dagger}$) is an annihilation (creation) operator of 
a boson at the lattice site $i$ with spin $\sigma=-1,0,\;{\rm or} \;1$, 
and $\langle i,j\rangle$ denotes the summation over nearest 
neighbor sites. We also introduce here the hopping integral $t$. We
absorb a constant single site occupation energy into the definition of $\mu$.
The effective parameter of our model 
$t$ can be derived from the microscopic details of the optical lattice
\cite{Jak05,Maz06}, assuming that the lattice 
site orbitals correspond to localized Wannier functions with one level
per site. We neglect the harmonic trap confinement in the following. 

The interaction part $H^{\rm int}$ is constructed under an assumption
that the total spin of the system is conserved and that the interaction amplitude 
is local \cite{Pet00,Ued10}. 
While the first requirement 
is natural due to general conservation laws,
the second assumption is justified for cold 
atoms due to their neutrality and short-range character of interacting forces. 
The total spin $S$ of two interacting spin-one bosons
attains three possible values $S=0,1,2$.
Because of the bosonic symmetry of the wave functions, in the
presence of the local interaction only $S=0$ and $S=2$ terms contribute
in $H^{\rm int}$. 
The resulting 
interaction amplitudes $U_S$ are proportional to the scattering lengths $a_S$ for each $S$ channel \cite{Jak05}.
Following Ref.\ \cite{Ima03} we write $H^{\rm int}$ in terms of the number operator
$n_i$ and spin operator ${\bf S}_i$ at the lattice site $i$:
\begin{equation}
H^{\rm int} = \frac{g_n}{2}\sum_in_i(n_i - 1) + \frac{g_s}{2} \sum_i(\mathbf{S}^2_i - 2 n_i), \label{hint1}
\end{equation}
were $g_n= (2U_2 + U_0)/3$ and $g_s= (U_2 - U_0)/3$.
Both the hopping integral $t$ and the
interaction strength $U_S$ can be tuned to
become of comparable magnitude by manipulating the laser light producing the
optical lattice.

In this paper we are interested in the effects of 
attractive interaction giving rise to pairing between spin-one bosons.
We introduce an auxiliary annihilation (and creation)
operator for a Cooper pair of bosons  
$B^{S,M}_{ij}= \sum_{\sigma\sigma'} C^{S,M}_{\sigma\sigma'}  b_{i \sigma}  b_{j \sigma'}$, 
where $C^{S,M}_{\sigma\sigma'}$ is the Clebsch-Gordan coefficient for the total spin $S$
and with spin projection $M$.
The explicit
form of the pair operators can be found in Ref. \cite{Ued10}.
The Hamiltonian \eqref{hint1} takes a new, compact form
\begin{eqnarray}
H^{\rm int}=\sum_i \left(U_0 B^{0,0\dagger}_{ii} B^{0,0}_{ii}+ U_2 \sum_{M=-2}^2 B^{2,M\dagger}_{ii} B^{2,M}_{ii} \right),\nonumber \\ \label{hint2}
\end{eqnarray}
which is more appropriate here since it  shows directly all structures of bosonic pair
correlations and hints to possible order parameters. 
In the next Section we solve the model \eqref{hubbard} with \eqref{kinetic} and \eqref{hint1}
within a Hartree-Fock mean-field approximation (MFA) and discuss
possible condensed phases of bosonic Cooper pairs.

\section{Mean-field approximation}
\label{solution}

The kinetic part \eqref{kinetic} of our model Hamiltonian 
is diagonal in the momentum representation
\begin{equation}
H^0=\sum_{k \sigma} \xi_k b_{k\sigma}^{\dagger} b_{k\sigma},
 \label{diag}
\end{equation}
where $b_{k\sigma}^+$ ($b_{k\sigma}$) is the creation (annihilation) operator for a 
particle with the lattice momentum ${\mathbf k}$ and the single particle kinetic energy
is denoted by $\xi_k$. We keep a general form of the dispersion relation $\xi_k$
in our derivation of the self-consistent equations and use a
specific model later in Section \ref{singlet}.

The construction of the appropriate mean-field Hamiltonian can 
be done within a textbook rule \cite{Bru04} 
by splitting two-body operators into 
paring operators and their
non-vanishing expectation values. This approximation consequently
neglects fluctuations. 
The form of our model Hamiltonian \eqref{hint2}
suggests the following choice for the pair expectation value 
\begin{equation}
\Lambda^{S,M}=\langle B^{S,M}_{ii}\rangle, 
\label{lambdaSM}
\end{equation}
which can also be expressed as
$\Lambda^{S,M}=\sum_{\sigma\sigma'} C^{S,M}_{\sigma\sigma'} \Lambda_{\sigma\sigma'}$,
with $\Lambda_{\sigma\sigma'}= \tfrac{1}{N_s} \sum_k \langle b_{k \sigma} b_{-k \sigma'}\rangle$.
The expectation values are taken at thermal equilibrium with the inverse temperature $\beta$ and
$N_s$ denotes the number of lattice sites.
We also allow for nonzero normal density expectation values by defining
an average site occupation matrix
$n_{\sigma\sigma'}= \tfrac{1}{N_s} \sum_k \langle b_{k \sigma}^\dagger b_{k \sigma'}\rangle$. Throughout the paper we deal with  quantities 
described by $3\times 3$ matrices in the spin index,
such as $n_{\sigma\sigma'}$ or $\Lambda_{\sigma\sigma'}$. 
Therefore, we introduce here a more compact matrix notation $\hat n$ and $\hat \Lambda$ for those quantities.

Within MFA \cite{Eva65,Bru04} the Hamiltonian \eqref{hint1} takes the following form 
\begin{equation}
H^{\mathrm{int}}_{\mathrm{MF}} = 
\sum_{k\sigma\sigma'}\left( b^\dagger_{k \sigma} w_{\sigma\sigma'} b_{k \sigma'} + 
\tfrac{1}{2}(b^\dagger_{k\sigma} \Delta_{\sigma\sigma'} b^\dagger_{-k\sigma'} +
\mathrm{h.c.})\right)-E_0N_s.\label{hintMF}
\end{equation}
In the above equation a matrix valued order parameter appears
\begin{equation}
\hat \Delta = U_0 \hat C^{0,0} \Lambda^{0,0} + U_2\sum^2_{M=-2} \hat C^{2,M} \Lambda^{2,M},
\label{deldef}
\end{equation}
which describes spontaneous symmetry breaking due to BCS-type paring. 
The quantity 
\begin{equation}
\hat w = 2 \left(U_0 \hat C^{0,0} \hat n \hat C^{0,0} + U_2 \sum^2_{M=-2} \hat C^{2,M} \hat n  \hat C^{2,M} \right)\label{wdef} 
\end{equation}
describes the effective Hartee-Fock potential. Note that the Clebsch coefficients
for fixed $S,M$ are also represented by a $3\times 3$ matrix $\hat C^{S,M}$.
We keep the additive constant
\begin{eqnarray}
E_0&=&
\tfrac{U_0 - U_2}{2}\left[2\mathrm{Tr}(\hat n^T\hat C^{0,0}\hat n\hat C^{0,0}) + 
|\mathrm{Tr}(\hat\Lambda \hat C^{0,0})|^2\right]\nonumber\\
&&+\tfrac{U_2}{2}\left[\mathrm{Tr} (\hat n^2 + \hat\Lambda^\dagger \hat\Lambda ) + n^2\right],
\label{ezero}
\end{eqnarray}
which is necessary in the discussion of phase stabilities
presented in the Section \ref{stability}.

We finally arrive at the mean field self-consistent equations by calculating
the normal $n_{\sigma\sigma'}$ and anomalous $\Lambda_{\sigma\sigma'}$ averages 
in the grand canonical ensemble with the quadratic interaction Hamiltonian \eqref{hintMF}. 
This procedure is equivalent \cite{Bru04} to the requirement of attaining a minimum 
of the free energy with the Hamiltonian \eqref{hintMF}, when $\hat \Delta$ and $\hat w$
are variational parameters. The technical details  
of the derivation are given in the Appendix \ref{diagapp}.
Here we present the final result obtained from \eqref{aver} and \eqref{averM}
\begin{equation}
\begin{pmatrix}
\openone + \hat n^* & -\hat \Lambda \\ 
\hat \Lambda^* & -\hat n  
\end{pmatrix}
= \tfrac{1}{N_s}\sum_{k}\left(1 - e^{-\beta M_k}\right)^{-1},
\label{selfcons}
\end{equation}
where $M_k$ is a Bogoliubov--de Gennes matrix
\begin{equation}
M_k =  \begin{pmatrix}
  (\xi_k-\mu) \openone + \hat w & \hat \Delta \\ -\hat \Delta^* & - (\xi_k-\mu) \openone - \hat  w^*
  \end{pmatrix}.
\label{defM}
\end{equation}

For the purpose of solving the self-consistency equations 
\eqref{selfcons} and \eqref{defM} in practice it is convenient to simplify the expression for
$\hat \Delta$ given in \eqref{deldef} and for $\hat w$ in \eqref{wdef}. With the help of general algebraic identities \cite{ident} applied to the matrices $\hat \Lambda$ and $\hat n$ we arrive at
\begin{equation}
\hat \Delta = (U_0-U_2)\hat C^{0,0}\mathrm{Tr}(\hat C^{0,0}\hat \Lambda)+U_2\hat \Lambda,
\label{delsimpl}
\end{equation}
\begin{equation}
\hat w = 2(U_0-U_2)\hat C^{0,0}\hat n\hat C^{0,0}+U_2(\hat n^T+n \hat \openone),
\label{wsimpl}
\end{equation}
where all $\hat C^{S,M}$ have been eliminated except of $\hat C^{0,0}$.

\section{Symmetry classification of ordered states}
\label{symmetry}

The accepted strategy, which allows to classify the solutions for the matrix order 
parameter from the self-consistent equations, relays on symmetry considerations \cite{Vol00}.
The symmetry arguments alone allow to identify stationary states of the free energy,
as it was done recently
for the spinor condensates \cite{Ho99,Yip07}. Here we need not only to identify
the symmetry classified states, but we also want to investigate the
phase diagram as a function of the interaction parameters.
Therefore, we have to compare
free energy of symmetry classified phases to find the minimal one.
In the investigation
of superfluid $^3$He it was observed \cite{Vol86}, but not strictly proven,
that the phase possessing the
highest remaining symmetry
corresponds indeed to a local, and very often to the global free energy minimum.

We follow the standard symmetry classification approach. 
We start by determining the highest allowed symmetry phase, and then we consider
the solutions with a lower symmetry. For the sake of completeness of
the presentation we give below a more detailed account of this derivation.
We will use the classification introduced in this Section  
to determine numerically the phase diagram,
by solving the non-linear mean field equations within
a given symmetry class. 

The full symmetry of our system (in a generic case $U_2\neq U_0$)
involves the gauge and the spin rotation symmetry, so it is $U(1)\times SO(3)$.
This symmetry is smaller  then in the superfluid $^3$He case, which has 
$U(1)\times SO(3)\times SO(3)$ symmetry group.
The possibility of breaking the gauge invariance is crucial
in our search of the phases with pairing. Symmetry of our
system allows
the gauge symmetry to be broken not only independently,
but also in a combination with the spin symmetry operation.
Thus we have to consider also the possibility of gauge-spin 
symmetry breaking, similar to superfluid $^3$He.

\subsection{Symmetry transformations}

We start the discussion of symmetry with the global $U(1)$ gauge symmetry transformation
$b_{k \sigma} \rightarrow e^{i\psi} b_{k \sigma}$,
where $\psi$ is a constant phase.
 The single site occupation
matrix $\hat n$ is gauge invariant, so from \eqref{wdef}
it follows that $\hat w$ is gauge invariant as well. The pair expectation
value transforms as $\hat \Lambda \rightarrow e^{2i\psi} \hat \Lambda$, which
substituted to \eqref{deldef}
leads to the order parameter transformation $\hat \Delta \rightarrow e^{2i\psi} \hat \Delta$.
It is easy to check that this gauge transformation is a symmetry of our mean field 
equation \eqref{selfcons} with \eqref{defM}. 

The spin rotation SO(3) is described by a unitary matrix $\hat r$, which acts as follows:
$b_{k\sigma} \rightarrow \sum_{\sigma'} r_{\sigma\sigma'} b_{k\sigma'}$. The general 
rotation matrix $\hat r$ can be parameterized by three Euler angles of elementary rotations
generated by three components of the spin-one operator. From the definitions
of the averages $\hat n$ and $\hat \Lambda$ we obtain the
transformation rules
\begin{equation}
\hat n \rightarrow \hat r^* \hat n \hat r^T,\qquad  
\hat \Lambda \rightarrow \hat r \hat \Lambda \hat r^T.
\label{rotn}
\end{equation}
The above transformations substituted to \eqref{wsimpl} and \eqref{delsimpl} 
give the following spin rotation of the effective potential 
and the pairing order parameter: 
\begin{equation}
\hat w \rightarrow \hat r \hat w \hat r^\dagger,\qquad 
\hat \Delta \hat \eta  \rightarrow \hat r  \hat \Delta \hat \eta \hat r^\dagger,
\label{rotDel}
\end{equation}
where $\hat \eta=\sqrt{3}\hat C^{0,0}$. We have used 
the identity $\hat \eta \hat r^* \hat \eta =\hat r$, which
follows from the explicit form $\eta_{\sigma\sigma'}=-(-1)^{\sigma}\delta_{\sigma,-\sigma'}$.
One can check that the right-hand side of the mean field
equation Eq.\ \eqref{selfcons} consequently transforms as
$M_k\rightarrow R M_k R^\dagger$, with 
a unitary $R=\mathrm{diag}(\hat r, \hat r^*)$, where $\mathrm{diag}$
stands for a block diagonal matrix.
The left-hand side of this equation transforms upon \eqref{rotn} in 
the same manner, thus verifying the $SO(3)$ spin rotation 
symmetry of our mean-field formulation.

\subsection{Continuous symmetry phases}
\label{chapcont}

{\em No broken symmetry.} The requirement of invariance upon the full symmetry 
transformation $U(1)\times SO(3)$ applied
to $\hat w$ and $\hat \Delta$ gives as the only solution 
$\hat w=w \hat \openone$ and $\hat \Delta=0$,
where $w=\frac{2}{9}(U_0+5U_2) n $. The single site density
is $n$ and the occupation matrix 
reads $\hat n=\frac{1}{3}n\hat \openone $.
This describes a free boson gas with a renormalized chemical
potential due to the Hartree-Fock treatment of the contact interaction.

{\em Singlet phase.} The highest possible symmetry phase with non-zero pairing amplitude
arises when we break the $U(1)$ gauge symmetry, but leave the 
spin rotation symmetry. We derive from the invariance condition
\begin{equation}
 \hat \Delta \hat \eta = \hat r \hat \Delta \hat \eta \hat r^\dagger
\end{equation}
that for a general $\hat r$ the order parameter has to be 
$\hat \Delta \hat \eta =\Delta \hat \openone $ with some complex $\Delta$ and
$\hat w = w \hat \openone$, with $w$ the same as in the free case discussed above. 
Going back to Eq.\ \eqref{deldef}
we find that the expectation values of the bosonic pair operators $\Lambda^{S,M}$
are non-zero only for $S=0$ in this $SO(3)$ symmetric phase. We will 
call this 
phase the \emph{singlet phase} as pairing happens only in the singlet channel,
with the finite order parameter $\hat \Delta = U_0\hat C^{0,0}\Lambda^{0,0}$.

{\em Quintet phase.} We search now for paired phases, which allow for
non-vanishing $S=2$ (i.e. quintet) components of the order parameter. The simplest way to achieve this is by 
lowering the  spin rotation $SO(3)$ symmetry to an axial $U(1)$ symmetry. 
We choose an arbitrary quantization axis and express
the spin rotations around this axis as $\hat r(\varphi)=e^{i\varphi \hat S_z} $, 
with some
angle $\varphi$ and $\hat S_z=\hat C^{2,2}- \hat C^{2,-2}$.
We require now a more general spin-gauge invariance
condition for the order parameter
\begin{equation}
 \hat \Delta \hat \eta = e^{2i\psi} \hat r(\varphi) \hat \Delta \hat \eta  \hat r(\varphi)^\dagger,
\label{spingauge}
\end{equation}
where the gauge symmetry breaking phase $\psi$ can now
depend on the spin rotation angle $\varphi$. We
obtain three different solutions, which are presented below: 
\begin{subequations}
\label{all}
\begin{align}
U(1)_{S_z-\varphi}:\quad \psi&=-\varphi & \hat \Delta &= \Delta \hat C^{2,2}, \label{phi}\\
U(1)_{S_z-\frac{\varphi}{2}}:\quad \psi&=-\varphi/2& \hat \Delta &= \Delta \hat C^{2,1}, \label{phi2}\\
U(1)_{S_z}:\quad \psi&=0& \hat \Delta &= \Delta \hat C^{0,0} + \Delta' \hat C^{2,0}.
\label{zero}
\end{align}
\end{subequations}
We follow the
notation of Ref.~\cite{Vol86} to label the above spin-rotation
breaking axial phases.
Remaining solutions with $+\varphi$, and $+\varphi/2$
can be obtained by changing the direction of the quantization
axis, so they do not describe a different symmetry phase.

The only non-zero pair expectation amplitude
is $\Lambda^{2,2}$ for the $U(1)_{S_z-\varphi}$ phase
and $\Lambda^{2,1}$ for $U(1)_{S_z-\frac{\varphi}{2}}$, which follows from 
the comparison of $\hat \Delta$ definition in \eqref{deldef} with the result \eqref{all}.
In these two axial phases 
the symmetry allows for pairing only in the \emph{quintet channel}. The remaining $U(1)_{S_z}$ phase
has a mixed singlet--quintet pairing order parameter, which has to be parameterized
by two (complex) numbers $\Delta$ and $\Delta'$.

The Hartree--Fock potential $\hat w$ in all the axial phases is 
restricted by the symmetry to
be diagonal. This brings a possibility of magnetic order,
coexisting with the pairing, 
marked by spin rotation symmetry breaking in the spin dependent site occupation.

\subsection{Discrete symmetry phase}

Within only $3\times 3$ 
matrix representations one cannot construct the icosahedral or octahedral
symmetry, without allowing for generation of all possible rotations.
The biggest non-trivial discrete symmetry is thus $T$ -- the symmetry group 
of tetrahedron without reflections. The 
set of group generators can be
explicitly expressed as $\{\openone, e^{i\frac{2\pi}{3} \hat S_z}, 
e^{i\pi(\hat S_z+\sqrt{2}\hat S_x)/\sqrt{3})}\}$, 
where we use $3\times 3$ matrix representation of spin one with 
$\hat S_x= \hat C^{2,1} + \hat C^{2,-1}$.
Substituting these generators for $\hat r$ in the
invariance condition \eqref{spingauge} we obtain as the only 
solution $\psi=\tfrac{2\pi}{3}$ and
$\hat \Delta=\Delta (\hat C^{2,2}+\sqrt{2} \hat C^{2,-1})$. 

\section{Mean-field solution for singlet phase}
\label{singlet}

The singlet phase, introduced from the symmetry arguments
in Section \ref{chapcont} is our natural candidate for a physically
attainable phase.
The system in the singlet phase has a maximal remaining
symmetry of all the phases with non-zero pairing. The singlet phase is
unitary, meaning that the matrix order parameter $\hat \Delta$ 
is proportional to a unitary matrix.
Stable phases of liquid $^3$He were previously found to be 
unitary \cite{Vol86} as well.

Simple form of the order parameter $\hat \Delta=\Delta \hat \eta$ 
in the singlet phase leads to an identity $M_k^2=e_k^2\openone_{6\times 6}$,
where the Bogoliubov--de Gennes matrix $M_k$ was defined in \eqref{defM}. 
The quasi-particle excitation energy
\begin{equation}
e_k=\sqrt{(\xi_k-\mu+w)^2-|\Delta|^2}
\label{spec}
\end{equation}
is a triple degenerate eigenvalue of $M_k$ as defined
in Eq.\ \eqref{diagM}. 
We can now  directly calculate the generalized occupation factor
in the mean field equation \eqref{selfcons}
\begin{equation}
\left(1 - e^{-\beta M_k}\right)^{-1}=f(e_k)M_k+\tfrac{1}{2},
\label{cosh}
\end{equation}
with $f(e_k)=\frac{\coth (\beta e_k/2)}{2 e_k}$. We recall that $M_k$
depends on $\hat w$ and $\hat \Delta$, which are related to $\hat n$ and $\hat \Lambda$:
\begin{subequations}
\label{w_del}
\begin{align}
\hat w&=\tfrac{2}{3}(U_0+5U_2)\hat n,\\
\hat \Delta&=U_0\hat \Lambda,
\end{align}
\end{subequations}
in the singlet phase, as obtained in section \ref{chapcont}. We substitute \eqref{w_del} into $M_k$ in 
\eqref{cosh} and then equate to
the left-hand side of Eq.\ \eqref{selfcons}. The resulting self-consistent
equations in the singlet phase take a simple form
\begin{subequations}
\label{selfbcs}
\begin{align}
\frac{n}{3}&= \tfrac{1}{N_s}  \sum_{k} \left( \sqrt{e_k^2+|\Delta|^2}f(e_k) - \tfrac{1}{2} \right), \\
-\frac{1}{U_0}&= \tfrac{1}{N_s} \sum_{k}f(e_k).
\end{align}
\end{subequations}
The first equation provides a relation between the average occupation $n$
and the chemical potential $\mu$, while the second 
guarantees a non-zero paring amplitude. The form of this second 
equation is similar to the gap equation in the BCS theory, but with
a different function $f(e_k)$ due to boson statistics of condensating
quasiparticles.

Interestingly, these equations are formally equivalent to the one obtained
in the case of scalar attracting bosons in Ref. \cite{Sto94}.  The only difference
is that the optical lattice provides a natural ultraviolet cutoff in our model. 

The existence of the BCS type singlet solutions depends only on the strength $U_0$ of attraction
in the singlet channel and is insensitive to scattering in the quintet
channel. We will show in the next Section that the singlet paired phase of attracting
bosons can be stabilized by a repulsive quintet interaction. This is in a
marked difference to the scalar case, where the system always undergoes
a mechanical collapse before reaching Evans-Rashid transition \cite{Sto94}.

We note that the quasiparticles in BCS type bosonic condensate may undergo 
a statistical (Bose-Einstein) condensation \cite{Eva65,Sto94}. The transition 
occurs when the excitation spectrum in Eq. \eqref{spec} becomes gapless \cite{Cao07}.
The singular condition $e_{k=0}=0$ can be satisfied in a thermodynamic limit
for
\begin{equation}
|\Delta|=\xi_{k=0}-\mu+w,
\label{const}
\end{equation}
which fixes the chemical potential similarly to a standard BEC.
We separate the $k=0$ terms to obtain
\begin{equation}
\tfrac{1}{N_s}\sum_k f(e_k) = \frac{n_{\text{BEC}}}{3|\Delta|}+
\tfrac{1}{N_s}\sum_{k\neq 0} f(e_k),
\label{decomp}
\end{equation}
where we introduce $n_{\text{BEC}}$ --
a finite average density for quasi-particles with $k=0$ only.
The self-consistent equations for the BEC quasiparticle phase follow
\begin{subequations}
\label{selfbec}
\begin{align}
\frac{n}{3}&= \frac{n_{\text{BEC}}}{3}+\tfrac{1}{N_s}  \sum_{k\neq 0} \left( \sqrt{e_k^2+|\Delta|^2}f(e_k) - \tfrac{1}{2} \right), \\
-\frac{1}{U_0}&= \frac{n_{\text{BEC}}}{3|\Delta|} + \tfrac{1}{N_s} \sum_{k\neq 0}f(e_k),
\end{align}
\end{subequations}
when we substitute the
decomposition \eqref{decomp} into \eqref{selfbcs}.
The chemical potential $\mu$ is fixed by \eqref{const}, so $n_\text{BEC}$
becomes a new thermodynamic parameter, which we have to determine. 
Therefore, we distinguish two different phases: i) BCS phase where $n_{\rm BEC}=0$ and $\Delta \neq 0$, 
and ii) BEC phase where $n_{\rm BEC}\neq 0$ and $\Delta \neq 0$. 
The condition for the BCS/BEC borderline is obviously $n_\text{BEC}\rightarrow 0$, it is when 
\eqref{selfbec} reduces to \eqref{selfbcs}.
We note that the transition between BCS and BEC in the boson case cannot be interpreted as being a counterpart of the BCS-BEC crossover known in the fermionic condensed systems \cite{Leggett80,Noz85}.

\begin{figure}[tb]
\centerline{%
\includegraphics[width= 0.85\linewidth]{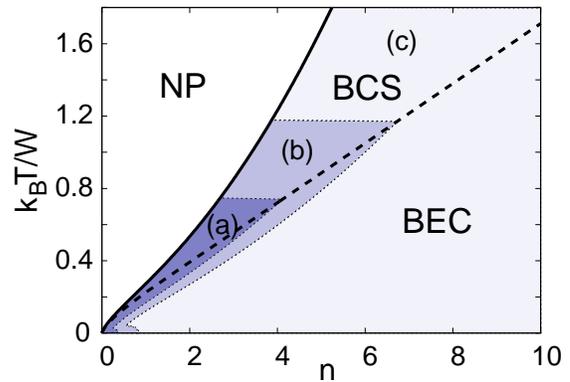}}%
\caption{The diagram of the singlet phase
presented  
in the density $n$ -- temperature $T$ coordinates at
$U_0 = -0.33W$.
The solid line denotes the boarder of BCS type 
phase with pairing (`NP' stands for no--pairing, normal phase), 
the dashed line marks the borderline of BEC quasiparticle 
condensate. 
For $U_2/|U_0|=0.59$ both phases in (c) (blank) region fulfill
the standard thermodynamic stability conditions,  
but are unstable in (a) (grey) and (b) (light-grey) regions.  For $U_2/|U_0|=0.64$
the regions (b) and (c) are thermodynamically stable, (a) region is unstable.
}
\label{phaseweak}
\end{figure}

\begin{figure}[tb]
\centerline{%
\includegraphics[width= 0.85\linewidth]{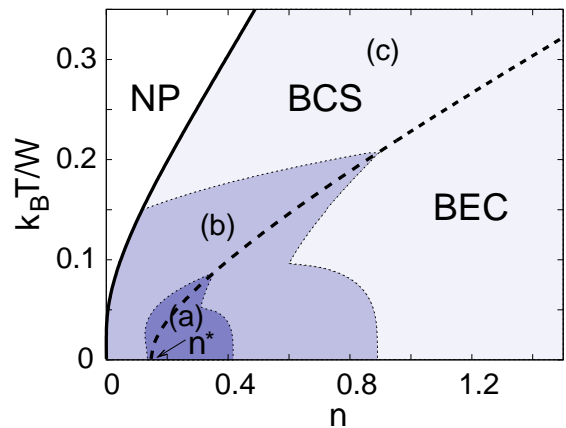}}%
\caption{The diagram of the singlet phase for strong interaction
$U_0 = -W$ in the density $n$ -- temperature $T$ coordinates.
The zero temperature
critical density is $n^*\approx 0.15$. 
The solid line denotes the border of BCS type 
paired phase, 
the dashed line marks the borderline of BEC quasiparticle 
condensate. Stability regions (a), (b) and (c) are 
defined in the same way as in Fig.\ \ref{phaseweak}. 
}
\label{phasestrong}
\end{figure}

In order to solve numerically Eqs.\ \eqref{selfbcs} and
\eqref{selfbec} 
we only need to provide the density of states 
$\rho(E)=\sum_k\delta(E-\xi_k)$
in the optical lattice.
We assume a simple elliptic model for the density
$\rho(E)=\frac{8N_s}{\pi W^2}\sqrt{(W/2)^2-E^2}$,
where $W=(32\pi)^{2/3}t$ is chosen to fit a low-energy density profile obtained
from the dispersion relation $\xi_k$. 

We are able to gain some analytical insight into the solution for this BEC singlet
phase. The integrals appearing in \eqref{selfbec} can be performed
in the limit
$T\rightarrow 0$ leading to
\begin{subequations}
\label{tzero}
\begin{align}
n &= n_{\text{BEC}}+ \tfrac{3}{\pi}\left[(1 - \omega^2)\arctan{\tfrac{1}{\sqrt{\omega}}} + \sqrt{\omega}(1+ \omega)\right] - \tfrac{3}{2},\\
-\frac{W}{U_0} &=  \frac{2n_{\text{BEC}}}{3\omega} + \tfrac{4}{\pi}\left[(1 + \omega)\arctan{ \tfrac{1}{\sqrt{\omega}} } - \sqrt{\omega}\right],
\label{omtzero}
\end{align}
\end{subequations}
where $\omega=\frac{2|\Delta|}{W}$. These two non-linear algebraic equations
determine $n_{\text{BEC}}$ and $|\Delta|/W$ for a given $n$. We define a critical
density $n^*$ by setting $n_{\text{BEC}}=0$ in \eqref{tzero}. The 
BEC condensate solution with
finite $n_{\text{BEC}}$ exists for densities larger than $n^*$ at $T=0$.
For weak interactions
$|U_0|/W<\frac{1}{2}$ we find only $n^*=0$ solution, 
which means that there is only the BEC phase at $T=0$.
For stronger interactions $|U_0|/W>\frac{1}{2}$ we find a region
of BCS phase extending down to zero temperature.
We present the resulting finite temperature phase diagram in the weak
interaction case
in Fig.\ \ref{phaseweak} for a fixed attractive interaction $U_0=-0.33W$.
The situation with strong interaction is illustrated in Fig.\ \ref{phasestrong} for $U_0=-W$.
Additionally, one can obtain an analytic solution $n^*=\frac{32}{3\pi^2}|U_0|/W$ 
for $|U_0|/W\gg 1$. 

\section{Stability}
\label{stability}

We discuss below the standard thermodynamic stability conditions expressed by: 
i) positivity of pressure
$p$, ii) positivity of constant volume specific heat $c_V$, and iii) 
positivity of isothermic compressibility
$\kappa_T$ (for the calculation see Appendix B). 

{\it Singlet phase.}
We find that $c_V$ is 
positive in the singlet phase and is
independent of $U_2$. The pressure $p$ and the 
inverse compressibility $\kappa_T^{-1}$ have a following linear dependence
on $U_2$:
\begin{subequations}
\label{pres}
\begin{align}
p(U_0,U_2)&=p(U_0)+U_2\tfrac{5n^2}{9a^3},\\
\kappa_T^{-1}(U_0,U_2)&=\kappa_T^{-1}(U_0)+ U_2\tfrac{10n^2}{9a^3}.
\end{align}
\end{subequations}
This means that for any point on the phase diagram in Fig.\ \ref{phaseweak}
or Fig.\ \ref{phasestrong}
we can find $U_2$ large enough to stabilize the BCS or BEC singlet phase.
The shadowed regions in these figures exemplify stability 
for $|U_2|/U_0=0.59$ and $0.64$, respectively.

\begin{figure}[tb]
\centerline{%
\includegraphics[width= 0.85\linewidth]{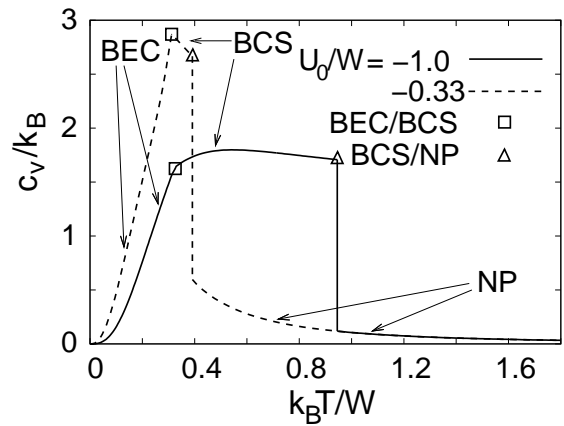}}%
\caption{The specific heat versus temperature for a fixed
density $n=1.5$. The solid line illustrates the sequence of
transitions for interaction strength
$U_0 = -W$, the dashed line is for $U_0=-0.33W$.
}
\label{cvfig}
\end{figure}

We illustrate the nature of subsequent transition by plotting
the specific heat $c_V$ in Fig. \ref{cvfig}. We show the specific
heat dependence on the temperature $T$ at fixed density $n=1.5$, 
which is representative both for strong and weak
attraction and contains all three phases: non-pairing, BCS-like
and the BEC quasiparticle condensate.
The plot does not depend on the strength $U_2$, provided
it is strong enough to stabilize the phases.
The specific heat exhibits a jump at the onset of pairing
(marked by a triangle in Fig. \ref{cvfig}),
indicating that the corresponding transition is of second order.
The transition to BEC quasiparticle condensate is contiunuos
with a cusp in the specific heat dependence (marked by a square),
the behaviour being known in the usual Bogoliubov theory of BEC condensation \cite{Sto99}.

\begin{figure}[tb]
\centerline{%
\includegraphics[width= 0.85\linewidth]{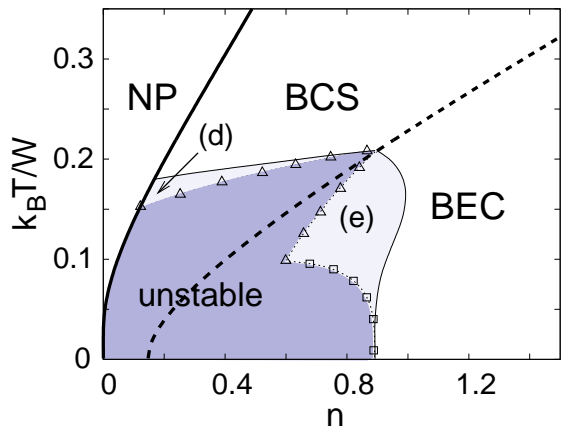}}%
\caption{
The phase diagram of the singlet phase with the same parameters as in Fig.~\ref{phasestrong}.
The binodal line is indicated by a thin solid curve. The system in both shadowed regions
(d) within BCS phase and (e) within BEC phase undergoes a spinodal decomposition.
The BCS phase above (d) is stable, same as BEC to the left of (e) region.
}
\label{spinodalfig}
\end{figure}

We inspect further the details of stability lines shown in Fig. \ref{phasestrong}.
We find generically two stable phases separated by an unstable
one for the system at fixed temperature. The system will then
have a tendency to spontaneously separate into the
dense and dilute phases. We make this statement qualitative
by considering the thermodynamic spinodal decomposition \cite{Spino95}
into the dense BEC and dilute BCS or normal phase.
The results are presented in Fig. \ref{spinodalfig}. The spinodal
stability lines are redrawn from Fig. \ref{phasestrong}, region (b). The line
marked by triangles is given by the compressibility condition
$\kappa_T^{-1}=0$, while the one marked by the squares
is given by the pressure $p=0$.
We find that the region denoted by light gray shadowing corresponds to
a metastable state, which undergoes the spinodal dcomposition. The
regions denoted BCS or BEC above the solid binodal line are stable
against such a thermal fluctuation.
This result indicates that the thermal fluctuations around the mean field solution
do not change qualitatively our phase diagram, they are of importance at the vicinity of the
stability borderlines. The role of quantum fluctuations is left
for future research.

We find that the most unstable point of our diagram is located at the
BCS/BEC borderline (marked by the thick dashed line in Fig. \ref{spinodalfig}).
We can solve the following conditions $p>0$ and $\partial p/\partial n>0$
at zero temperature on this corssover line, corresponding to 
the density $n^*$ (compare Fig. \ref{phasestrong}). We thus find $U_2^c$ -- the
critical strength of repulsion in the quintet channel at which the whole
BCS and BEC phases become stable. In the weak interaction regime
\begin{equation}
U_2^c=\frac{|U_0|}{10}\left(2+\frac{3}{1-|U_0|/2W}\right),
\end{equation}
which is valid for $|U_0|/W<\frac{1}{2}$, while in the strong interaction
regime
\begin{equation}
U_2^c=\frac{|U_0|}{20}\left(1+\frac{3\,\mathrm{arctan}(1/\sqrt{\omega})}{\sqrt{\omega}-\omega\mathrm{arctan}(1/\sqrt{\omega})}\right),
\end{equation}
valid for $|U_0|/W>\frac{1}{2}$ and with $\omega$ calculated from Eq.~\eqref{omtzero}
at $n_{\mathrm{BEC}}=0$. For the interaction strength $U_2>U_2^c$ there is
always a non-collapsing phase, which can be either normal,
BCS or BEC homogeneous, or an inhomogeneous mixture of the dilute and dense phases.

{\it Other symetry phases.} We have carried out a detailed, both analytical and numerical study of
stability \cite{Pel10} for the other phases. We find that both magnetic phases $U(1)_{S_z-\varphi}$
and $U(1)_{S_z-\varphi/2}$ are mechanically unstable. The tetrahedral T-phase with
the quintet pairing occurs for $U_2<0$ and $U_0>0$.
It can be made thermodynamically stable by increasing the 
singlet repulsion $U_0$. We find however, that the ferromagnetic
phase $U(1)_{S_z-\varphi}$ has lower free energy in the parameter regions, 
where T-phase becomes stable. Morover, we find that an infinitezimal
$\epsilon>0$ distortion of T-phase order paremeter by a magnetic contribution
$\Delta((1+\epsilon)\hat C^{2,2}+\sqrt{2}\hat C^{2,-1})$ leads to a
lower free energy. We conclude that the T-phase is not even metastable, as it
always corresponds to a saddle point of free energy.

\section{Conclusions and outlook}
\label{conclusions}

In summary, we applied the Hartree-Fock mean-field approximation to solve a problem of pairing between bosons with spin $S=1$ moving on  optical lattices. The order parameter describing such paired bosons has a matrix form. Detailed classiffiation of possible solutions according to their symmetries was presented. In particular, we found that the self-consistent equations for the $SO(3)$ symmetric phase have the same form as those for the scalar bosons. We showed that the coherent BCS type phase of paired bosons induced by attractive interaction in the singlet channel is stable provided that the interaction in a quintet channel is repulsive. This finding might be usefull in experiments to stabilize bosons with attractive interaction against mechanical collapse. 

The analyzed problem might be extended in the future in different ways. For example, it would be interesting to include local quantum correlation beyond the static Hartree-Fock approximation by using the bosonic dynamical mean-field theory developed recently \cite{Byczuk08}. Another line of research is to investigate inhomogeneous excited states of bosons with $S=1$ in the BCS or BEC phases, i.e. there should be generalized vortex states in the condensed phases because of the high remaining symmetry in the system. In the boson gas with hyper spin $S=2$, where a very reach variety of spinor BEC for repulsive interactions have been proposed \cite{Die06}, we expect stabilization of at least some of many symmetry allowed phases for attractive interactions.

\acknowledgments

The work of KB is supported by the grant N N202 103138 of Polish Ministry of Science and Education and, in part, by the grant the TRR80 of the Deutsche Forschungsgemeinschaft.

\appendix

\section{Diagonalization of the mean field Hamiltonian}
\label{diagapp}

In this section we diagonalize our mean field Hamiltonian \eqref{hintMF}
and derive Eq.\ \eqref{selfcons}. We use a convenient notation \cite{Sto99}, known 
as the Nambu notation in the standard theory of superconductivity for 
fermions \cite{Bru04} 
\begin{equation}
\boldsymbol\Phi_k = (b_{k1},\, b_{k0},\, b_{k-1},\, b_{-k1}^\dagger,\, b_{-k0}^\dagger,\,  b_{k-1}^\dagger)^T,\label{nambu}
\end{equation}
where the superscript $T$ denotes transposition.
The bosonic canonical commutation relations rewritten with the Nambu spinor are
$ [(\boldsymbol\Phi_{k})_\alpha, (\boldsymbol\Phi_{k'}^\dagger)_{\beta}]= 
(\Sigma_z)_{\alpha \beta}\delta_{k,k'}$ 
where $\alpha,\beta=1,\dots ,6$ and 
$\Sigma_z={\mathrm{diag}} (\hat \openone,-\hat \openone)$. Here $\Sigma_z$ is a diagonal 
matrix, the symbol $\hat \openone$ denotes the unity $3\times 3$ matrix.
We express
the model Hamiltonian in the mean field approximation \eqref{hintMF} as follows
\begin{equation} 
H_{\text{MF}}=H^0+H^{\text{int}}_{\text{MF}}-\mu N = \tfrac{1}{2} \sum_{k} 
 \boldsymbol\Phi^\dagger_{k} \Sigma_z M_k \boldsymbol\Phi_{k}
 + \mathrm{const.}.
\label{MFdef}
\end{equation}
The additive constant does not enter the calculation presented in this Appendix.
We introduce here the $6\times 6$ matrix $M_k$ resulting from the commutation 
$[\boldsymbol\Phi_k,H_{\mathrm MF}]=M_k\boldsymbol\Phi_k$. 
The explicit form of $M_k$ is given in Eq.\ \eqref{defM}.

With the Nambu spinor we write a compact expression for all normal and 
anomalous averages
\begin{equation} 
\tfrac{1}{N_s}\sum_{\vec k}\left\langle\boldsymbol\Phi_{k} \boldsymbol\Phi_{k}^\dagger\right\rangle =
\begin{pmatrix}
\hat \openone + \hat n^* & \hat \Lambda \\ 
\hat \Lambda^* & \hat n
\end{pmatrix},\label{aver} 
\end{equation}
where the single site occupation $\hat n$ and 
the amplitude of Cooper pair condensate $\hat \Lambda$ were defined in Section \ref{solution}.
Our aim is now to compute the l.h.s. of the above equation. This is easily done with
a suitable Bogoliubov transformation performed on the mean field Hamiltonian \eqref{hintMF}.
We follow this route by introducing a new Nambu spinor  $\boldsymbol\Gamma$
for quasiparticle excitations
\begin{equation}
\boldsymbol\Gamma_k = 
\begin{pmatrix}
\hat u_k & \hat v_k \\ 
\hat v^*_k & \hat u^*_k
\end{pmatrix}
\boldsymbol\Phi_{k},
\label{trans} 
\end{equation}
where $\hat u_k$ and $\hat v_k$ contain coefficients to be determined below. We require 
the new spinor $\boldsymbol\Gamma$ to describe proper quasiparticles, so 
$[\boldsymbol\Gamma_k,H_{\text MF}]=E_k\boldsymbol\Gamma_k$,
where $E_k={\mathrm{diag}}(\hat e_k, -\hat e_k)$ and 
$(\hat e_k)_{\sigma\sigma'}=e_{k\sigma}\delta_{\sigma\sigma'}$. The eigenvalues 
$e_{k\sigma}$ 
correspond  to the excitation energy of quasiparticles in the BCS condensate.
The components of $\boldsymbol\Gamma_k$ have to fulfill the bosonic 
commutation relations, namely
$ [(\boldsymbol\Gamma_{k})_\alpha, (\boldsymbol\Gamma_{k'}^\dagger)_{\beta}]= 
(\Sigma_z)_{\alpha \beta}\delta_{k,k'}$.
The corresponding requirement for the coefficients of the Bogoliubov transformation \eqref{trans} leads to
\begin{equation}
\begin{pmatrix}
\hat u_k & \hat v_k \\ 
\hat v^*_k & \hat u^*_k
\end{pmatrix}^{-1} 
= \Sigma_z 
\begin{pmatrix}
\hat u_k & \hat v_k \\ 
\hat v^*_k & \hat u^*_k
\end{pmatrix}^\dagger 
\Sigma_z.
\label{condition}
\end{equation}
Finally, we write the 
eigenvalue equation for the $6\times 6$ matrix $M_k$
\begin{equation}
\begin{pmatrix}
\hat  u_k &\hat  v_k \\ 
\hat v^*_k & \hat u^*_k
\end{pmatrix} 
M_k 
\begin{pmatrix} 
\hat u_k & \hat v_k \\ 
\hat v^*_k &\hat u^*_k
\end{pmatrix}^{-1} 
= E_k.\label{diagM}
\end{equation}
It follows from the hermicity of $\Sigma_z M_k$ and the transformation
constrain \eqref{condition} that
the eigenvalues of $M_k$ have to be real.
Note also that the additional matrix $\Sigma_z$ 
enters our derivation due to the bosonic commutation relation,
but is absent in the standard BCS formulation for the fermions. 

It turns out that we don't need to calculate explicitly
$\hat u_k$ and $\hat v_k$ from Eq.\ \eqref{diagM} as long as we 
are only interested in the thermodynamic averages.  
The quasiparticle averages are particularly simple
\begin{equation}
\left\langle \boldsymbol\Gamma_{k} 
\boldsymbol\Gamma_{k'}^\dagger\right\rangle 
\Sigma_z
= (1 - e^{-\beta E_k})^{-1}\delta_{k k'}.\label{averGamma}
\end{equation}
We transform this equation back to the original spinor
$\boldsymbol\Phi_k$ with the help of Eq.\ \eqref{trans} and we get
\begin{equation} 
\left\langle\boldsymbol\Phi_k 
\boldsymbol\Phi_k^\dagger\right\rangle
\Sigma_z = 
\begin{pmatrix} 
\hat u_k & \hat v_k \\ 
\hat v^*_k & \hat u^*_k
\end{pmatrix}^{-1}
 (1 - e^{-\beta E_k})^{-1} 
\begin{pmatrix}
 \hat u_k & \hat v_k \\ 
\hat v^*_k & \hat u^*_k
\end{pmatrix}, \label{averPhi}
\end{equation}
which simplifies to
\begin{equation} 
\left\langle\boldsymbol\Phi_k 
\boldsymbol\Phi_k^\dagger\right\rangle
\Sigma_z =  (1 - e^{-\beta M_k})^{-1}. 
\label{averM}
\end{equation}
The above equation together with Eq.\ \eqref{aver} gives the final result
of this Appendix.

\section{Calculation of thermodynamic parameters}
\label{thermapp}

We calculate the grand canonical potential (per site)
\begin{equation}
\Omega = -\tfrac{k_BT}{N_s}\ln \mathrm{Tr} e^{-\beta H_{\text{MF}}}
\end{equation}
by taking the trace $\mathrm{Tr}$ over all many-particle states of second-quantized 
$H_{\text{MF}}$ as
defined in \eqref{MFdef}.
The general expression for the entropy per site $S=-\frac{\partial \Omega}{\partial T}|_{\mu}$ 
is then
\begin{equation}
S=\frac{k_B}{N_s}(\ln \mathrm{Tr} e^{-\beta H_{\text{MF}}}+\beta \left<H_{\text{MF}}\right>_{\text{MF}})
\end{equation}
where $\left<\ldots\right>_{\text{MF}}$ denotes the thermodynamic average with our 
model mean field hamiltonian. Using the results of diagonalization of bilinear $H_{\text{MF}}$
derived
in Appendix A
we get
\begin{equation}
S=\tfrac{k_B}{N_s}\sum_{k\sigma}\left(-\ln(1-e^{-\beta e_{k\sigma}})+
\frac{\beta e_{k\sigma}}{e^{\beta e_{k\sigma}}-1}  \right),
\end{equation}
while the constant introduced in \eqref{hintMF} does not enter this expression.
With the above explicit of $S$ we write the final compact
formula for $\Omega$
\begin{equation}
\Omega = -TS+\tfrac{1}{N_s}\sum_{k}(\xi_k-\mu)n_{k}+E_0,
\end{equation}
where $n_k=\sqrt{e_k^2+|\Delta|^2}f(e_k)-\tfrac{1}{2}$ is average
occupation of a quasi-momentum state $k$. The constant has been
now properly recovered and is entirely included in $E_0$ as
defined in \eqref{ezero}. For the singlet phase there is a simple
scalar expression $E_0=\frac{3\Delta^2}{2U_0}+\frac{5U_2+U_0}{9}n^2$.
With the grand canonical potential $\Omega$ expressed entirely
in terms of the order parameter one calculates the 
pressure $p=-\Omega/a^3$, the specific heat per site 
$c_V=T\frac{\partial S}{\partial T}|_{n}$ and the inverse
compressibility $\kappa_T^{-1}=n\frac{\partial p}{\partial n}|_{T}$.




\begin{thebibliography}{99}
\bibitem{ident} The algebraic identities
$\sum_{S,M} \hat C^{S,M} \mathrm{Tr} (\hat C^{S,M} \hat A)^T=\hat A$
and 
$\sum_{S,M} \hat C^{S,M} \hat A (\hat C^{S,M})^T=  (\mathrm{Tr} \hat A) \hat \openone$,
follow from the completness of total spin $\left|S,M\right\rangle$ basis. 
They are a matrix form of the ortogonality condition 
for the Clebsch-Gordan coefficients.
\end{thebibliography}
\end{document}